
\input harvmac
\input amssym.def
\input epsf
\noblackbox
\newcount\figno
\figno=0
\def\fig#1#2#3{
\par\begingroup\parindent=0pt\leftskip=1cm\rightskip=1cm\parindent=0pt
\baselineskip=11pt
\global\advance\figno by 1
\midinsert
\epsfxsize=#3
\centerline{\epsfbox{#2}}
\vskip 12pt
\centerline{{\bf Figure \the\figno} #1}\par
\endinsert\endgroup\par}
\def\figlabel#1{\xdef#1{\the\figno}}
\def\pano{\par\noindent}

\def\pmb#1{\setbox0=\hbox{#1}%
 \kern-.025em\copy0\kern-\wd0
 \kern.05em\copy0\kern-\wd0
 \kern-.025em\raise.0433em\box0 }
\font\cmss=cmss10
\font\cmsss=cmss10 at 7pt
\def\inbar{\,\vrule height1.5ex width.4pt depth0pt}

\def\rlx{\relax\leavevmode}
\def\Cop{\relax\,\hbox{$\inbar\kern-.3em{\rm C}$}}
\def\Rop{\relax{\rm I\kern-.18em R}}
\def\Nop{\relax{\rm I\kern-.18em N}}
\def\one{\relax{\rm 1\kern-.25em I}}
\def\Pop{\relax{\rm I\kern-.18em P}}
\def\Zop{\rlx\leavevmode\ifmmode\mathchoice{\hbox{\cmss Z\kern-.4em Z}}
 {\hbox{\cmss Z\kern-.4em Z}}{\lower.9pt\hbox{\cmsss Z\kern-.36em Z}}
 {\lower1.2pt\hbox{\cmsss Z\kern-.36em Z}}\else{\cmss Z\kern-.4em
 Z}\fi}

\def\Dirac{\rlx\leavevmode \hbox{{$\cal D$} \kern-.99em /}\,}


\def\H{{\cal H}}

\def\ie{{\it i.e.}}
\def\eg{{\it e.g.}}
\def\cf{{\it cf.}}

\def\sl2r{SL_2\Rop}

\def\p{{\bf p}}
\def\oo{{\frak o}}
\def\so{{\frak s}\oo}

\def\sohat{\widehat{\so}}

\def\g{{\frak g}}

\def\h{{\frak h}}
\def\p{{\frak p}}

\def\Rcal{{\cal R}}

\def\End{\hbox{End}}

\def\ghat{\hat{\g}}
\def\hhat{\hat{\h}}

\def\Spin{{\cal S}}
\def\bV{{\cal V}}
\def\tK{\,^\tau K}




\lref\atiyahhirzebruch{
M.~F.~Atiyah, F.~Hirzebruch, {\it Vector bundles and homogeneous
spaces}, Proc. Sympos. Pure Math., Vol. III, 7 (1961).
}

\lref\helgason{
S.~Helgason, {\it Differential Geometry and Symmetric Spaces},
Academic Press, New York and London, 1962.
}

\lref\KS{Y.~Kazama, H.~Suzuki,
{\it New N=2 superconformal field theories and superstring compactification},
Nucl.\ Phys.\ B {\bf 321}, 232 (1989).}

\lref\minami{
H.~Minami, {\it K-groups of symmetric spaces, I/II}, Osaka
J. Math. {\bf 12}, 623 (1975); Osaka J. Math. {\bf 13}, 271 (1976).
}

\lref\Wittensymp{
S.~Axelrod, S.~Della Pietra, E.~Witten,
{\it Geometric quantization of Chern-Simons gauge theory},
J.\ Diff.\ Geom.\  {\bf 33}, 787 (1991).}

\lref\Wittengauged{
E.~Witten, {\it The N matrix model and gauged WZW models},
Nucl.\ Phys.\ B {\bf 371}, 191 (1992).}

\lref\etkpaper{
S.~Schafer-Nameki,
{\it D-branes in N = 2 coset models and twisted equivariant K-theory};
{\tt hep-th/0308058}.
}

\lref\FHTtwo{D.S. Freed, {\it The Verlinde algebra is twisted
equivariant K-theory}, Turkish J. Math. {\bf 25} no. 1, 159 (2001);
{\tt math.RT/0101038}.}

\lref\FHTcomplex{D.S. Freed, M.J. Hopkins, C. Teleman, {\it Twisted
equivariant K-theory with complex coefficients}, {\tt math.AT/0206257}.}      

\lref\FHTloop{D.S. Freed, M.J. Hopkins, C. Teleman, {\it 
Twisted K-theory and loop group representations I}; {\tt math.AT/0312155}.}

\lref\harveymoore{
J.~A.~Harvey, G.~W.~Moore,
{\it Algebras, BPS states, and strings},
Nucl.\ Phys.\ B {\bf 463}, 315 (1996);
{\tt hep-th/9510182}.}

\lref\bpsalgebra{
J.~A.~Harvey, G.~W.~Moore,
{\it On the algebras of BPS states},
Commun.\ Math.\ Phys.\  {\bf 197}, 489 (1998);
{\tt hep-th/9609017}.}

\lref\landweber{
~D.~Landweber , {\it
Multiplets of representations and Kostant's Dirac operator for equal
rank loop groups}, Duke Math. J. {\bf 110} (2001), no. 1, 121.}

\lref\Gawedzki{
K.~Gawedzki,
{\it Abelian and non-Abelian branes in WZW models and gerbes},
{\tt hep-th/0406072}.}

\lref\kostant{
B.~Kostant, {\it
A cubic Dirac operator and the emergence of Euler number multiplets of
representations for equal rank subgroups},  Duke Math. J. {\bf 100}
(1999), no. 3, 447.}

\lref\sloane{
{\tt http://www.research.att.com/$\tilde{\,}$njas/sequences/index.html};
The On-Line Encyclopedia of Integer Sequences.
}

\lref\MooreMinasian{
R.~Minasian and G.~W.~Moore, {\it K-theory and Ramond-Ramond charge},
JHEP {\bf 9711}, 002 (1997); {\tt hep-th/9710230}.}

\lref\WittenK{
E.~Witten, {\it D-branes and K-theory},
JHEP {\bf 9812}, 019 (1998); {\tt hep-th/9810188}.
}

\lref\MooreSegal{
G.~Moore, {\it Lectures on branes, K-theory and RR charges}, Clay
Mathematical Institute Lectures; {\tt http://www.physics.rutgers.edu/~gmoore/clay.html}.
}

\lref\Moorephysical{
G.~Moore,
{\it K-theory from a physical perspective},
{\tt hep-th/0304018}.}

\lref\mrgnonbps{
M.~R.~Gaberdiel, {\it Lectures on non-BPS Dirichlet branes},
Class.\ Quant.\ Grav.\  {\bf 17}, 3483 (2000); {\tt hep-th/0005029}.}

\lref\FSWZW{
S.~Fredenhagen, V.~Schomerus,
{\it Branes on group manifolds, gluon condensates, and twisted K-theory},
JHEP {\bf 0104}, 007 (2001);
{\tt hep-th/0012164}.
}

\lref\MMSII{
J.~Maldacena, G.~Moore, N.~Seiberg, {\it D-brane instantons and
K-theory charges}; {\tt hep-th/0108100}.
}

\lref\BSN{
V.~Braun, S.~Schafer-Nameki,
{\it Supersymmetric WZW models and twisted K-theory of SO(3)};
{\tt hep-th/0403287}.}

\lref\volkerK{V. Braun, {\it  Twisted K-theory of Lie groups}; {\tt hep-th/0305178}. }

\lref\GKO{P. Goddard, A. Kent, D. Olive, {\it Unitary representations
 of the Virasoro and supervirasoro algebras}, Commun.\ Math.\ Phys.\
 {\bf 103}, 105 (1986).}

\lref\gepnerFI{
D.~Gepner, {\it Field identification in coset conformal field theories},
Phys.\ Lett.\ B {\bf 222}, 207 (1989).}

\lref\horivafa{
K.~Hori, A.~Iqbal, C.~Vafa, {\it D-branes and mirror symmetry}; {\tt hep-th/0005247}.
}

\lref\MMSI{
J.~Maldacena, G.~Moore, N.~Seiberg, {\it Geometrical interpretation
of D-branes in gauged WZW models}, JHEP {\bf 0107}, 046 (2001); {\tt hep-th/0105038}.}

\lref\LW{W.~Lerche, J.~Walcher, {\it Boundary rings and N = 2 coset models},
Nucl.\ Phys.\ B {\bf 625}, 97 (2002); {\tt hep-th/0011107}.}

\lref\Mickelsson{
J.~Mickelsson,
{\it Gerbes, (twisted) K-theory, and the supersymmetric WZW model},
{\tt hep-th/0206139}.
}

\lref\vbssntocome{
V.~Braun, S.~Schafer-Nameki, in preparation.
}

\lref\stefantocome{
S.~Fredenhagen, {\it D-brane charges and boundary RG flows in coset models}, in preparation.
}

\lref\stefanBF{
S.~Fredenhagen,
{\it Organizing boundary RG flows}, Nucl.\ Phys.\ B {\bf 660}, 436
(2003); {\tt hep-th/0301229}.}

\lref\GaGatwist{
M.~R.~Gaberdiel, T.~Gannon,
{\it The charges of a twisted brane},
JHEP {\bf 0401}, 018 (2004);
{\tt hep-th/0311242}.}

\lref\GaGaI{
M.~R.~Gaberdiel, T.~Gannon,
{\it D-brane charges on non-simply connected groups},
JHEP {\bf 0404}, 030 (2004);
{\tt hep-th/0403011}.
}

\lref\GaGaRoI{
M.~R.~Gaberdiel, T.~Gannon, D.~Roggenkamp,
{\it The D-branes of SU(n)},
{\tt hep-th/0403271}.
}

\lref\StefanSO{
S.~Fredenhagen,
{\it D-brane charges on SO(3)},
{\tt hep-th/0404017}.
}

\lref\GaGaRoII{
M.~R.~Gaberdiel, T.~Gannon, D.~Roggenkamp,
{\it The coset D-branes of SU(n)},
{\tt hep-th/0404112}.} 

\lref\douglasfiol{
M.~R.~Douglas, B.~Fiol, {\it D-branes and discrete torsion. II}, {\tt hep-th/9903031}.}

\lref\BDLR{I.~Brunner, M.~R.~Douglas, A.~E.~Lawrence, C.~R\"omelsberger,
{\it D-branes on the quintic}, JHEP {\bf 0008}, 015 (2000); {\tt
hep-th/9906200}.}

\lref\horitorsion{
K.~Hori,
{\it Boundary RG flows of N=2 minimal models},
{\tt hep-th/0401139}.
}

\lref\LVW{W.~Lerche, C.~Vafa, N.~P.~Warner, {\it Chiral rings in N=2
 superconformal theories}, Nucl.\ Phys.\ B {\bf 324}, 427 (1989).}

\lref\FS{S.~Fredenhagen, V.~Schomerus,
{\it D-branes in coset models}, JHEP {\bf 0202}, 005 (2002); {\tt
hep-th/0111189}. }

\lref\SegalKG{G. Segal, {\it Equivariant K-theory}, Inst. Hautes
\'Etudes Sci. Publ. Math. {\bf 34}, 129 (1968).}

\lref\atiyahsegal{
M.F. Atiyah, G.B. Segal, 
{\it Equivariant K-theory, and completion}, 
J. Differential Geometry {\bf 3}, 1 (1969).}

\lref\CBMS{J.~P.~May, {\it Equivariant Homotopy and Cohomology
Theory}, CBMS, Regional Conference Series {\bf 91}, AMS, 1996.}

\lref\atiyah{
M.F. Atiyah, {\it K-theory Past and Present}, {\tt math.KT/0012213}}

\lref\freedloops{
D.~S.~Freed, {\it Twisted K-theory and loop groups};
{\tt math.at/0206237}.}

\lref\GaGa{
M.~R.~Gaberdiel, T.~Gannon,
{\it D-brane charges on non-simply connected groups},
{\tt hep-th/0403011}.}

\lref\AtiyahHop{
M.~Atiyah, M.~Hopkins,
{\it A Variant of K-theory: $K_{+-}$},
{\tt math.kt/0302128}.}

\lref\wittenK{
E.~Witten,
{\it D-branes and K-theory},
JHEP {\bf 9812}, 019 (1998);
{\tt hep-th/9810188}.
}

\lref\gao{
M.~R.~Gaberdiel, S.~Schafer-Nameki,
{\it Non-BPS D-branes and M-theory},
JHEP {\bf 0109}, 028 (2001);
{\tt hep-th/0108202}.
}

\lref\kapustinli{
A.~Kapustin, Y.~Li,
{\it D-branes in topological minimal models: The Landau-Ginzburg approach};
{\tt hep-th/0306001}.
}

\lref\hodgkin{L. Hodgkin, {\it The equivariant K\"unneth theorem in
K-theorem}, in {\it Topics in K-theory. Two independent
contributions}, Lecture Notes in Math. {\bf 496}, Springer, Berlin, 1975. }

\lref\snaith{V.P. Snaith, {\it On the K\"unneth formula spectral
sequence in equivariant $K$-theory}, Proc. Cambridge Philos. Soc. {\bf
72}, 167, (1972).}

\lref\pittie{H.V. Pittie, {\it Homogeneous vector bundles on
homogeneous spaces}, Topology {\bf 11}, 199 (1972).}

\lref\gepnerfusion{
D.~Gepner, {\it Fusion rings and geometry},
Commun.\ Math.\ Phys.\  {\bf 141}, 381 (1991).
}

\lref\Halpern{
M.~B.~Halpern, E.~Kiritsis, N.~A.~Obers, K.~Clubok,
{\it Irrational conformal field theory},
Phys.\ Rept.\  {\bf 265}, 1 (1996);
{\tt hep-th/9501144}.}

\lref\PS{A. Pressley, G. Segal, {\it Loop Groups}, Oxford (1986).}

\lref\FH{W.~Fulton, J.~Harris, {\it Representation Theory}, Springer NYC,
GTM (1991)}

\lref\adams{J.F. Adams, {\it Vector Fields on spheres}, Ann. od
 Math. {\bf 74}, 603 (1962)}

\lref\BlauTh{
M.~Blau, F.~Hussain, G.~Thompson,
{\it Grassmannian Topological Kazama-Suzuki Models and Cohomology},
Nucl.\ Phys.\ B {\bf 488}, 599 (1997);
{\tt hep-th/9510194}.}

\lref\WittenGrassmann{E. Witten, {\it The Verlinde Algebra And The
Cohomology Of The Grassmannian}, {\tt hep-th/9312104}.}



\Title{\vbox{
\hbox{hep-th/0408060}
\hbox{DESY-04-063}}}
{\vbox{\centerline{K-theoretical boundary rings in ${\cal N}=2$ coset models}}}
\smallskip
\smallskip
\bigskip
\centerline{
Sakura Sch\"afer-Nameki\ \footnote{$^{\sharp}$}{{\tt e-mail: 
S.Schafer-Nameki@damtp.cam.ac.uk }}
\footnote{$\,$}{{\tt $\qquad\qquad$ Sakura.Schafer-Nameki@desy.de}}
}
\bigskip
\centerline{\it II. Institut f\"ur Theoretische Physik,
University of Hamburg}  
\centerline{\it Luruper Chaussee 149, 22761 Hamburg, Germany}
\smallskip
\vskip2cm
\centerline{\bf Abstract}
\bigskip

\noindent
A boundary ring for ${\cal N}=2$ coset conformal field theories is
defined in terms of a twisted equivariant K-theory. The
twisted equivariant K-theories ${}^\tau K_H(G)$ for compact Lie groups $(G, H)$ such that $G/H$ is
hermitian symmetric are computed. These turn out to have the same ranks as the ${\cal
N}=2$ chiral rings of the associated coset conformal field theories,
however the product structure differs from that on chiral
primaries. In view of the K-theory classification of D-brane charges this suggests an
interpretation of the twisted K-theory as a `boundary
ring'. Complementing this, the ${\cal N}=2$ chiral ring is studied in
view of the isomorphism between the Verlinde algebra $V_k(G)$ and 
${}^\tau K_G(G)$ as proven by Freed, Hopkins and Teleman. 
As a spin-off, we provide explicit formulae for the ranks of the
Verlinde algebras. 

\bigskip

\Date{8.~8.~2004}


\newsec{Introduction}

Twisted K-theory has recently attracted much attention in various areas in
string theory and conformal field theory. The two main applications
that have crystalized so far are the classification of
D-brane charges in backgrounds with non-trivial NSNS 3-form flux, as
well as the beautiful result by Freed, Hopkins and
Teleman (FHT) \refs{\FHTcomplex,\FHTtwo, \FHTloop}, which identifies the
Verlinde algebra for a compact Lie group $G$ with the twisted
equivariant K-theory $\tK_{G}(G)$. The main motivation for the present
work was to set out and ask, whether there is a similar
relation between twisted equivariant K-theory
and the ${\cal N}=2$ chiral ring for coset conformal field
theories. For coset models with groups $(G,H)$ and common centre $Z$,
the relevant twisted equivariant K-theories are
$\tK_{H/Z}(G)$. 
We compute these K-theories in general, thus extending the results of \refs{\etkpaper}, and show them to be of
the same rank as the ${\cal N}=2$ chiral ring of the coset model in question. However, interestingly,
the product defined on the K-theory
differs from the one on chiral primaries. This  
new ring-structure will be referred to as the {\it (K-theoretical)
boundary ring}, as it has a natural interpretation in terms of
a product structure on (classes of) D-branes. This
may have a close correspondence with the algebra of BPS states defined
in \refs{\harveymoore, \bpsalgebra}. A CFT-discussion of this product
will appear in \refs{\stefantocome}, extending some of \refs{\LW}.

In the context of D-branes in supersymmetric WZW and coset theories 
the identification of D-brane charges with twisted K-theory has been established
in various instances. The case of WZW  models and the corresponding
computations of twisted K-theories for compact
Lie groups is discussed in \refs{\FSWZW, \MMSII, \Moorephysical,
\volkerK,\GaGatwist,\GaGaI, \BSN, \StefanSO, 
\GaGaRoI , \GaGaRoII, \Gawedzki}. For some of
the ${\cal N}=2$ coset models, namely the Grassmannian cosets,
the charge lattices for the D-branes were obtained in \refs{\MMSI,
\LW, \horivafa, \douglasfiol, \BDLR, \FS} and the relevant twisted, equivariant
K-theories have been computed in \refs{\etkpaper}. One task, which will
be accomplished in the present paper is to generalize the computation
in \refs{\etkpaper} to all ${\cal N}=2$ Kazama-Suzuki coset models \refs{\KS}.
Despite the successful description of D-brane charges in these
theories by means of twisted K-theory, a conceptual understanding of
this relation still needs to be elucidated. Some progress to this end has
been obtained for topological theories in \refs{\MooreSegal}. 

On the other hand, the theorem by FHT \refs{\FHTcomplex, \FHTtwo,
\FHTloop} provides a concrete correspondence between conformal field
theoretical data, such as the Verlinde fusion ring, and topology. From
a CFT (or rather TFT) point of view, the result by FHT on $\tK_G(G)$ can be
interpreted as a statement about the D-brane charges in the $G/G$
gauged WZW model, which is in fact topological. The next
simplest such theories are the Kazama-Suzuki models -- which are
conformal, but have ${\cal N}=2$ worldsheet supersymmetry, and thus
would allow for a topological twisting. The present paper provides a discussion of these ${\cal N}=2$
coset models in light of the results in \refs{\FHTcomplex, \FHTtwo,
\FHTloop}. 
In summary we shall prove the following
\smallskip
\noindent {\bf Theorem.} {\it 
Let $G$ be a simple, simply-connected,
connected Lie group and $H$ a connected, maximal rank subgroup of $G$,
such that $G/H$ is hermitian symmetric.
Let $Z$ be the common centre of $G$ and $H$, which is assumed to act
without fixed points, and denote by ${\cal
R}_{cp}^{(G,H)}$ the chiral ring of the corresponding ${\cal N}=2$
coset conformal field theory.
Then
\eqn\mainresult{
\hbox{rank}\left(\tK_{H/Z}^{\dim(G)} (G)\right) \cong 
\hbox{rank}\left({\cal R}^{(G,H)}_{cp}\right) \,,
}
and the ring structure on the K-theory is 
\eqn\theoremone{
\tK_{H/Z}^{\dim(G)} (G) \cong \left({R_H \over I_k(G)}\right)^Z \,.
}
Here, $\tK_{H/Z}^{\dim(G)} (G)$ is the twisted $H/Z$-equivariant K-theory of
$G$, where the action of $H$ on $G$ is by conjugation, $R_H$ denotes the
$H$-representation ring, $I_k(G)$ is the Verlinde ideal of $G$ and the
$Z$-invariant part is taken on the RHS. The twisting $\tau\in H^3_{H}(G)$ is
related to the level of the coset model by $\tau = \kappa [{\bf H}]$, where
$[{\bf H}]$ is the generator of $H^3_H(G)$ and $\kappa = k+ g^\vee$.  
}
\smallskip
\noindent
Assuming the K-theory classification of D-brane charges
\refs{\MooreMinasian, \WittenK}, a straight forward implication of the
theorem is the following 
\smallskip
\noindent {\bf Corollary.} {\it 
The charge lattice for D-branes in the Kazama-Suzuki coset models
associated to $(G,H)$ is of the same rank as the ${\cal N}=2$ chiral ring.
}
\smallskip
\noindent
As emphasized, the ring structure on the K-theory is however somewhat
different from the one on the chiral ring, thus motivating the 
\smallskip
\noindent 
{\bf Definition.} 
{\it The (K-theoretical) boundary ring ${\cal B}_k^{(G,H)}$ of the ${\cal N}=2$ coset model is defined as the ring in \theoremone.}
\smallskip
\noindent
The plan of this paper is as follows. In section 2, we present our
main result by computing 
the twisted equivariant K-theories relevant for all Kazama-Suzuki (KS) cosets,
generalizing \refs{\etkpaper}, and prove that the
ranks of the K-theory agrees in all instances with that of the chiral
ring. We provide the explicit formulae for the ranks, including the ranks of
the Verlinde algebras (which to our knowledge have not been explicitly
documented in the literature), the computation of which we provide in
appendix A. A new boundary ring, motivated by K-theory, is
defined in section 3, and its relation to the standard (bulk) chiral
ring is discussed.
The construction of elements of the chiral ring as K-theory
classes using families of affine Dirac operators is provided in
section 4 and we close in section 5 with discussions and outlook.


\newsec{Twisted equivariant K-theory for Kazama-Suzuki models}

GKO coset models \refs{\GKO} (see also \refs{\Halpern} for further references) with ${\cal N}=1$ supersymmetry associated to a compact Lie group $G$ and a maximal rank
subgroup $H$, with corresponding Lie algebras $\g$ and $\h$,
have chiral algebra 
\eqn\cosetmodel{
{\cal A} =  {\ghat_{k} \oplus \sohat (\dim (\g/\h))_1\over
\hat{\h}_{k+h^\vee_\g - h^\vee_\h }} \,,
}
and are known to be ${\cal N}=2$ supersymmetric
if the right coset space $G/H$ is a hermitian symmetric space \refs{\KS}. 
The so-obtained Kazama-Suzuki (KS) coset models are thus classified by the
(irreducible) hermitian symmetric spaces, which fall into the
following classes \refs{\helgason, \KS}:

\vskip4pt
$$
\def\tbntry#1{\vbox to 23 pt{\vfill \hbox{#1}\vfill }}
\hbox{\vrule 
      \vbox{\hrule 
            \hbox{\vrule
                  \hbox to 80 pt{
                  \hfill\tbntry{$ G$}\hfill }
                  \vrule 
                  \hbox to 120 pt{
                  \hfill\tbntry{ $H $ }\hfill }
                  \vrule 
                }
            \hrule 
            \hbox{\vrule
                  \hbox to 80 pt{
                  \hfill\tbntry{$ SU(n+m)$}\hfill }
                  \vrule 
                  \hbox to 120 pt{
                  \hfill\tbntry{ $SU(n) \times SU(m) \times U(1)$ }\hfill }
                  \vrule 
                }          
            \hrule 
            \hbox{\vrule
                  \hbox to 80 pt{
                  \hfill\tbntry{$SO(n+2) $}\hfill }
                  \vrule 
                  \hbox to 120 pt{
                  \hfill\tbntry{ $SO(n)\times SO(2) $ }\hfill }
                  \vrule 
                }
            \hrule 
            \hbox{\vrule
                  \hbox to 80 pt{
                  \hfill\tbntry{$SO(2n) $}\hfill }
                  \vrule 
                  \hbox to 120 pt{
                  \hfill\tbntry{ $SU(n)\times U(1)$ }\hfill }
                  \vrule 
                }          
            \hrule 
            \hbox{\vrule
                  \hbox to 80 pt{
                  \hfill\tbntry{$ Sp(2n)$}\hfill }
                  \vrule 
                  \hbox to 120 pt{
                  \hfill\tbntry{ $ SU(n)\times U(1)$ }\hfill }
                  \vrule 
                }
            \hrule 
            \hbox{\vrule
                  \hbox to 80 pt{
                  \hfill\tbntry{$E_6 $}\hfill }
                  \vrule 
                  \hbox to 120 pt{
                  \hfill\tbntry{ $SO(10)\times U(1) $ }\hfill }
                  \vrule 
                }         
            \hrule 
            \hbox{\vrule
                  \hbox to 80 pt{
                  \hfill\tbntry{$E_7 $}\hfill }
                  \vrule 
                  \hbox to 120 pt{
                  \hfill\tbntry{ $E_6\times U(1) $ }\hfill }
                  \vrule 
                }   
            \hrule     
        }
     }
$$
\centerline{
\hbox{{\bf Table 1:} {\it Hermitian Symmetric Spaces.}}} 
\vskip 8pt
Recall, that $G/H$ is hermitian symmetric iff the following condition
is satisfied on the Lie algebras: consider the orthogonal
decomposition $\g=\h \oplus \p$,
with $[\h, \h]\subset \h$, then the
condition reads 
\eqn\hermsymcond{
[\h, \p]\subset\p \,,\qquad 
[\p, \p] =0 \,,  
}
\ie, in particular $\p$ is abelian. More generally the coset is
K\"ahler if $[\p, \p]\subset \p$. Note, that under these
circumstances, the maximal tori of $G$ and $H$ can be chosen to coincide. 

In the following we shall be interested in the corresponding twisted
equivariant K-theories ${}^\tau K_{H}(G)$. The K-theories in the case
of projective cosets $SU(n+1)/U(n)$ and generalized superparafermions $SU(n+1)/U(1)^n$
have been computed in \refs{\etkpaper}. 
The computation there relied on the group $G$ being a connected,
simply-connected Lie group. The first step in order to generalize
these K-theory computations to all the coset models corresponding to
the spaces listed in table 1, is to note that for hermitian symmetric spaces $G/H$ \helgason (Theorem 4.6)
\eqn\coveriso{
G/H \cong \widetilde{G}/\widetilde{H} \,,
}
where $\widetilde{G}$ denotes the covering group of $G$. In particular,
the cosets based on $SO(n)$ can be replaced by the corresponding
$Spin(n)$ cosets, and thus \coveriso\ allows to treat the K-theory
computation uniformly for all KS models, assuming that all the groups
are simply-connected.

In the following we shall always assume that $G$ is a simple, simply-connected, connected Lie group,
and $H$ a connected, maximal rank subgroup of $G$.
Under these circumstances, we have shown in \refs{\etkpaper} 
that the twisted equivariant
K-theories can be computed using the observation that
\eqn\etkobserve{
{}^\tau K_H(G) = {}^\tau K_G(G \times_H G^L)\,,
}
where $G^L$ is acted upon by left-multiplication, whereas the action
on the remaining groups is by conjugation. This yields by the
equivariant K\"unneth theorem
\eqn\etkresult{
{}^\tau K_H (G) 
= {}^\tau K_G(G) \otimes_{R_G} R_H 
= {R_G\over I_k(G)} \otimes_{R_G} R_H \,,
}
where we invoked the result of Freed, Hopkins and
Teleman \refs{\FHTtwo,\FHTloop}
\eqn\fhtresult{
{}^\tau K_G(G) = V_k(G) = {R_G\over I_k(G)} \,.
}
$V_k$ denotes the Verlinde algebra\footnote{$^\natural$}{As
coefficients in $\Zop$ are used, the correct terminology is in fact {\it ring}
instead of {\it algebra}. However, we shall continue to refer to this
as the {\it Verlinde algebra}, making the coefficient ring/field
explicit, when necessary.} 
and $I_k(G)$ the Verlinde ideal of $G$
at level $k$, which is specified by the twisting $\tau$. $H$ acts
upon $G$ by the conjugation action. 
Note that \fhtresult\ is in fact an algebra isomorphism, where the product on
$V_k$ is the fusion product and on the twisted K-theory side it is the
Pontryagin product \refs{\FHTcomplex, \FHTtwo, \FHTloop}. We shall
discuss the induced product structure on ${}^\tau K_{H}(G)$ in the
next section.
Further, since $H$ is connected and of maximal rank, $R_H$ is free as
an $R_G$-module \refs{\pittie}, so that
\eqn\etkcont{
{}^\tau K_H (G) 
= {R_H\over I_k(G)} \,.
}
Thus, in order to determine the rank of the K-theory, we need to
compute the rank of $R_H$ as an $R_G$-module, as well as the rank
of the Verlinde algebra. In order to acquire the former, we recall
that by \refs{\pittie}
\eqn\repringrank{
K(G/H) = R_{H} \otimes_{R_G} \Zop\,,
}
\ie, the rank of $R_H$ can be computed via the
untwisted K-theory of the symmetric spaces (left-action cosets) in question. By a theorem
of Atiyah and Hirzebruch \refs{\atiyahhirzebruch} (Theorem 3.6)
\eqn\ah{
K(G/H) = \Zop^{|W_G|\over |W_H|}\,,
}
where $W_G$ denotes the Weyl group of $G$ (a more detailed
discussion of these K-groups can be found in \refs{\minami}).
Thus we arrive at the final result
\eqn\etkfinal{
{}^\tau K_H (G) = \Zop^{d_k(G)  {|W_G|\over |W_H|}}\,,
}
where 
\eqn\dGk{
d_k(G) = \hbox{rank} (V_k(G)) \,.
}
In order to acquire a totally explicit expression for the rank, we need to determine
$d_k(G)$ in each of the above cases. One can \eg\ determine $d_k(G)$
combinatorially. $d_k(G)$ is equal to the number of integrable
highest weights at level $k$, \ie, it can be determined as the number
of solutions to the inequality
\eqn\inthw{
(\Lambda, \theta) \leq k \,,
} 
where $\Lambda$ denotes the highest weight, and $\theta$ the highest
root. That is, one has to count the non-negative integer solutions for
the Dynkin labels $\{\Lambda^{(i)}\}$ respecting the inequality
\eqn\tosolve{
\sum_{i=1}^n \Lambda^{(i)} a_i^{\vee } \leq k \,, 
}
where $a_i^{\vee }$ are the dual Coxeter labels. 
Doing the combinatorics, the details of which we provide in appendix
A, implies table 2 in appendix A. 

Due to the non-trivial selection rules in the KS coset theories, the
relevant K-groups that should classify the D-brane charges are in fact
${}^\tau K_{H/Z} (G)$, where $Z$ is the common centre of $G$ and
$H$. We shall restrict our attention to the cases, when $Z$ acts
without fixed points. 
As explained in \refs{\etkpaper}, this reduces the rank of the
K-theory by a factor equal to the lengths, $l(Z)$, of the orbits of $Z$ acting
on the geometric invariant theoretical (GIT) quotient $H/\!/H$, so that
\eqn\etkks{
{}^\tau K_{H/Z} (G) = \Zop^{{d_k(G)\over l(Z)}\,   {|W_G|\over |W_H|}}\,.
}
Note that the charge lattice is thus precisely of the same rank as the
chiral ring of the KS models as determined in \LVW. 


\newsec{${\cal N}=2$ boundary rings from K-theory}

In view of the result \etkks\ it is very tempting to conjecture that the
chiral ring of an ${\cal N}=2$ coset model is given by a
twisted, equivariant K-theory -- much like the Verlinde algebra is
$\tK_G(G)$. In this section we will discuss this correspondence in
some detail, and arrive at the conclusion that the twisted K-theory
defines a ring, whose underlying $\Zop$-module structure is the same as the
chiral ring (\ie, they have the same ranks), but the product structure
is different.


\subsec{Proposal for an ${\cal N}=2$ boundary ring}

The K-theory for the KS-coset models naturally comes equipped with a
product structure. Tracing this through our computations in the last
section, we see that this is the induced ring structure from
$\tK_G(G)$, which by FHT is the Pontryagin product on K-theory
classes\footnote{$^{\flat}$}{Note that this makes use of
the product on $G$ in an essential way.} and agrees with the fusion
product in the Verlinde algebra of $G$.
 
Let us assume the validity of the conjectural one-to-one correspondence between
K-theory classes and classes of D-branes (where the equivalence is say
with respect to boundary RG flows).
Put into this context, our K-theory computation suggests to define the 
following (K-theoretical) boundary ring 
\eqn\Kring{
{\cal B}_{k}^{(G,H)}
:= \tK_{H/Z}^{\dim (G)}(G) \cong \left(V_k(G) \otimes_{R_G} R_H \right)^Z\,,
}
where the $Z$-invariant part is taken on the RHS. We shall mostly abbreviate
this as ${\cal B}_k$.
Let us stress, that this is different from the chiral ring of the
coset model. 
In particular, \Kring\ can be written as a quotient of the
$H$-representation ring by the Verlinde ideal of $G$ 
\eqn\boundaryring{
{\cal B}_k \cong  \left({R_H \over I_k(G)}\right)^Z \,.
}
The D-brane interpretation of this is twofold: firstly, the K-theory charge
lattice seems to be spanned already by the Cardy branes (labeled by chiral
primaries). This is presumably due to the worldsheet ${\cal N}=2$ supersymmetry.
An interesting exercise, which might elucidate this point is to analyze
the charges in the topologically twisted Kazama-Suzuki
models. The second point is, that the K-theory comes naturally with a product
structure, which therefore corresponds to a product on
equivalence classes of D-branes. Having said this, the ring structure should
then in particular account for the charge
relations, that can be derived \eg\ from a worldsheet point of view. 
A complementary CFT discussion of this matter will appear in
\refs{\stefantocome}. For the $SU(2)/U(1)$ coset model, the next
section will illustrate that the
ring structure does indeed respect the charge relations derived in
\refs{\MMSI}. 


\subsec{${\cal N}=2$ super-minimal models}

The simplest KS models are the super-minimal models/superparafermions,
realized in terms of $SU(2)/U(1)$.
Recall that
\eqn\sutwoone{\eqalign{
\tK_{U(1)}(SU(2)) &= {R_{u(1)} \over I_k(SU(2))}
= {\Zop[\zeta, \zeta^{-1}]\over
\langle \hbox{Sym}_{k+1}(\zeta +\zeta^{-1})=0 \rangle} \cr
&=\left\langle 1, \zeta,
\zeta^{-1}, \cdots\,;\ \hbox{Sym}_{k+1}(\zeta+\zeta^{-1})= 0 \right\rangle
 \,.
}}
Here, $\hbox{Sym}_{n}(x)$ denotes the symmetric polynomial of degree
$n$ in $x$ and the generator, $\Lambda$, of $R_{SU(2)}$ has been decomposed with
respect to $U(1)$, \ie, $\Lambda =(\zeta + \zeta^{-1})$. 
For the boundary ring one needs to consider $\tK_{U(1)/Z}(SU(2))$,
\ie\ take the invariant part under the common centre $Z=\pm 1$, which
acts on the representations as $\zeta\mapsto -\zeta$,
thus removing the odd powers of $\zeta$. Hence
\eqn\boundarymini{
{\cal B}_k = \left\langle 1, \zeta^2,
\zeta^{-2}, \cdots ;\ \hbox{Sym}_{k+1}(\zeta+\zeta^{-1})=0 \right\rangle
 \,.
}
In particular, the rank is $k+1$ and does indeed agree with the one of
the chiral ring. However, the relations in the latter are $\Lambda^{k+1}=0$,
whereas they are $\hbox{Sym}_{k+1}(\zeta +\zeta^{-1})=0$ in the K-theory, so that the ring
structures differ\footnote{$^\sharp$}{I thank S. Fredenhagen
for discussions on this point.}. For instance at $k=1$ the
relation reads $\zeta^2 + 1 + \zeta^{-2}=0$, which does not factor
within the ring ${\cal B}_{k}$. 

Note that if one considers only the 
homogeneous part in $\Lambda$ of the Verlinde ideal, that is in this case the
ideal $J= \langle (\zeta+\zeta^{-1})^{k+1}=0\rangle$, the resulting
ring would agree with the chiral ring. 
{\it E.g.} for $k=1$, the relation is $\zeta^2 + 2+ \zeta^{-2} =0$, which generates the same ideal as $\zeta^4 + 2 \zeta^2 + 1= (\zeta^2+1)^2 =0$. Thus,
setting $x=\zeta^2 +1$, we obtain the same relation as in the chiral
ring. Let us stress that this is however {\it not} what one obtains
from the K-theory and thus in ${\cal B}_k$. For level 1, the latter is
the quantum deformation of the chiral ring (which is simply
$H^\ast(\Cop P^1)$) with the deformation parameter set to $-1$. This
observation has in fact been made in \refs{\LW} for the ring obtained
from boundary intersection matrices of D-branes in KS models. 
 
Note also, that the boundary ring nicely encodes the charge relations in the ${\cal N}=2$ super-minimal models. Geometrically the
(A-)branes are lines in the disc target space \refs{\MMSI}. The shortest
lines correspond to the basis of the charge lattice, with the relation
that the closed ring of shortest branes is trivial
\refs{\MMSI,\stefanBF}. This is precisely the relation in ${\cal
B}_k$, under the identification of the short branes with the generators $\zeta^{l}$, $l\in 2
\Nop$. 

\noindent
{\it Remark}

\noindent
We should digress, and make a remark upon the relation of our results
for the super-minimal models to the recent
computations in \refs{\horitorsion}. The computation of D-brane
charges in the ${\cal N}=2$ minimal models $su(2)_k/u(1)$ in the paper
in question yielded B-brane charges $\Zop_{k+2}$ and as well as
A-brane charges $\Zop^{k+1}$. This is not in contradiction with the present
results and the ones in \refs{\etkpaper}, as the computation in the
latter is for the diagonal modular invariant, whereas the computation
in \refs{\horitorsion} seems to be for a $(-1)^F$ orbifold thereof
(see also comments on this matter in \refs{\kapustinli}). A detailed discussion of this point will
appear in a forthcoming paper \refs{\vbssntocome}. In brief, for modular
invariants, which are obtained as simple-current extensions of the
diagonal modular invariant, one has to incorporate the additional
equivariance with respect to the simple-current in the corresponding
K-theory computation. For non-trivial actions on the fermions (\ie, on
the $\so(2d)_1$ factor), one has additional twist choices apart from
$H^3(X)$, and for instance the Hopkins K-groups $K_{\pm}$
\refs{\AtiyahHop} (see also \refs{\WittenK, \gao}) are relevant. An
example of this has been worked out in \refs{\BSN}.


\subsec{Boundary ring versus chiral ring: Level $1$ discussion}

In \refs{\LVW, \gepnerfusion} a geometrical interpretation of the chiral ring has
been given for level $1$ KS models, based on simply-laced
groups. There it was proven that
\eqn\lvwlevelone{
{\cal R}_{k=1}^{(G,H)} \cong H^\ast (G/H) \,, 
}
where the RHS is the cohomology ring for the right-action coset space. Let's
see what our proposal yields in this instance. Our twisted K-theory
computation results in
\eqn\ssnlevelone{
\tK_{H} (G) \cong V_{k=1}(G) \otimes_{R_{G}} K(G/H) \,.
}
Further note that for simply-laced groups, $|Z|= |V_{k=1}(G)|$. Taking
the $Z$-equivariance into account, we infer that 
\eqn\leveloneranks{
\hbox{rank} ({\cal R}_{k=1}^{(G,H)})= \hbox{ rank} (\tK_{H/Z} (G))  \,.
}
However the product structure on the chiral ring, which
in this case is the wedge product on the cohomology ring of $G/H$,
differs from the one on
\eqn\levelonefinished{
\tK_{H} (G) \cong \left({R_H \over I_1(G)}\right)^Z \,.
}
The reason is again that $Z$ does not have a homogeneous action on
the generators of $V_1(G)$.
Again, one sees that taking just the highest degree term in the
relations for the Verlinde ideal at level $1$ would give rise to the
chiral ring.


\subsec{Boundary ring versus chiral ring: Relation to fusion rings}

It was observed in \refs{\LVW, \gepnerfusion}, that the chiral
ring of an ${\cal N}=2$ coset model can be related to Verlinde
algebras (\ie, fusion rings). This is most concisely explained by
Witten in \refs{\Wittengauged}.  
The chiral ring is obtained by quantizing the following phase space
\eqn\phasespace{
P_R= {T \times T/Z \over W_H} \,,
}
where $T$ is the maximal torus of $G$ (and so also of $H$). 

On the other hand one can relate the chiral ring to the representation
ring of $H$ by noting that \refs{\Wittensymp} the Verlinde
algebra for $H$ is obtained by quantizing the space
\eqn\phaseV{
P_V= {T\times T \over W_H}\,.
}
This yields the key relation, that after quantization we obtain
\eqn\keypoint{
{\cal R}_{cp}^{(G,H)} = \left(V_{\tilde{k}}(H)\right)^{Z} \,,
}
so that the chiral ring of the coset is (in fact only as a lattice) isomorphic to the $Z$-invariant
part of a quotient of $R_H$ by an ideal, which is not the Verlinde
ideal of $G$.
However, this line of argument is to be
taken with a grain of salt, as \keypoint\ only seems to hold as
a lattice isomorphism, one cannot infer straight away that the product
on the RHS of \keypoint\ is different from the chiral ring product,
\eg\ by considering simple examples.


\subsec{Product structure}

Let us briefly discuss the product on the
K-theory, without making use of the relation to $R_H$. 
One can define the Pontryagin product on
the K-theory classes, yet again, in complete analogy to
\refs{\FHTcomplex}, namely 
\eqn\ponpro{
m_{pon}\, :\qquad \tK_{H/Z}(G) \otimes \tK_{H/Z} (G) \ \rightarrow\  
\tK_{H/Z} (G) \,.}
To establish this, consider the multiplication on the group $m:\,
G\times G \rightarrow G$ and assume that the twisting respects this,
in the sense that $m^\ast (\tau)$ factorizes over the two groups. Then
by pushing forward along $m$ induces the product 
\eqn\pushforward{
\tK_{H/Z}(G)\,  \otimes_{R_G} \tK_{H/Z} (G) \ \rightarrow\ 
{}^{\tau \oplus \tau} K_{H/Z\times H/Z}(G\times G) \ \rightarrow\
{}^{\tau } K_{H/Z }(G\times G) \,,
}
where the first map is application of the K\"unneth theorem. This is
equivalent to the product, that we encountered in section 2, which we
obtained by invoking the product on $\tK_G(G)$ of FHT, \ie, the product
on \Kring. 
On the other hand the product on the chiral ring represented in terms of coset fields is
the fusion product induced from the Verlinde algebras of $G$ and
$H$. More precisely, it is the fusion product on pairs of primaries in
$V_k(G)$ and $V_{k+g^\vee}(H)$, respectively, modulo selection and
identification rules. In particular this is distinct from \Kring.


\newsec{The ${\cal N}=2$ chiral ring and twisted equivariant K-theory}

Recall that the FHT theorem states the isomorphism
\eqn\FHTstatement{
\tK^{\dim (G)}_G(G) \cong V_{k}(G) \,,
}
where $\tau = k+h^\vee$ is the twist-class in $H^3_G(G)$.
The proof of this theorem in \refs{\FHTloop} is based on constructing
the K-theory classes from families of affine Dirac operators on loop
space. In this section we divert from the main thread of the paper and
explain how a related construction can be put forward for the chiral
ring. Only at the end, we shall comment again on the relation to the
boundary ring. 

As a motivation, first recall the construction of the chiral ring in
\refs{\LVW} for the ${\cal N}=2$ coset theory with chiral algebra \cosetmodel.
The essential ingredient is an index computation for the
${\cal N}=2$ supercharges (affine Dirac operators), more precisely,
the chiral primaries (or, by spectral flow, the RR ground states) are solutions to
\eqn\rrgroundstates{
G^{\pm} |\, \varphi\, \rangle = 0\,,
}
where the affine Dirac operators are defined
as (neglecting for the present discussion irrelevant prefactors)
\eqn\affineDirac{
G^{\pm} = \Dirac_{L\ghat/L\hhat}^{\pm} 
= 
\left(
J^{\mp \alpha}_{-n} \psi^{\pm\alpha}_{n} - {1\over 12} \, f^{\pm
\gamma}_{\pm\alpha\, \pm\beta}\, \psi^{\pm\alpha}_n \,
\psi^{\pm\beta}_m\, 
\psi^{\mp\gamma}_{-n-m}\right) \,,
}
where $\psi^\alpha$ are adjoint fermions transforming under the $so(2d)_1$
algebra, and the $\alpha$'s are summed over the positive roots of $\g/\h$.
Note that for trivial $\h$, $\Dirac^+$ is precisely the affine Dirac operator that enters the
construction of the $G/G$ twisted K-theory in \refs{\FHTloop}.
In fact, the zeroes of the
affine Dirac operator $\Dirac^+$ 
for both the WZW and coset models have been computed by Landweber in
\refs{\landweber}, in analogy to the analysis for finite-dimensional Lie
groups by Kostant \refs{\kostant}. 

The key idea of Mickelsson
\refs{\Mickelsson} and Freed-Hopkins-Teleman \refs{\FHTloop}, which
provides the link to twisted K-theory, is to couple the Dirac-operator on
$L\g$ to an $L\g^\ast$-valued gauge field. The kernel of the affine
Dirac operator on $L\g$ is trivial, however the gauge-coupled Dirac
operator (in the following referred to as the FHT Dirac family) has
non-trivial kernel, to which one can associate a twisted K-theory
class on $G$.


\subsec{Review of Dirac-family construction for the Verlinde algebra}

First we review the Dirac-family construction for \FHTstatement.
Throughout this section we shall adopt the conventions of \refs{\FHTloop}.
Consider the following (gauge-coupled) family of affine Dirac operators
\eqn\MickD{
\Dirac^{FHT}_{\mu} = \Dirac_{L\g} + i \psi(\mu) \,,
}
for $\mu\in L\g^\ast$ (\cf\ (4.11) in \refs{\Mickelsson}, (11.2) in
\refs{\FHTloop}), where the affine Dirac operator on $L\g$ is defined by
\eqn\DiracLg{
\Dirac_{L\g} =  \sum_{\alpha\in \Delta_{L\g} }\left(
J^{\alpha}_{-n} \psi^{-\alpha}_{n} -{1\over 12} \sum_{ \beta, \gamma \in \Delta_{L\g}} f^{
\gamma}_{\alpha\, \beta}\, \psi^{\alpha}_n \,
\psi^{\beta}_m\, 
\psi^{-\gamma}_{-n-m}\right) \,.
}
Furthermore, $\psi (\mu)$ denotes Clifford multiplication with this
element of Cliff$(L\g)$. 
$\Dirac$ acts on $\H_\lambda \otimes \Spin_{L\g}$, where $\H_\lambda$
is an integrable highest weight representation of $\widehat{L\g}$, and
$\Spin$ the spin representation. 
The key properties of this family are:
The Dirac family \DiracLg\ is equivariant under
the co-adjoint action of $LG$. This can be mapped to the gauge action
of $LG$ on the space ${\frak A}$ of $\g$-valued connections on the circle by
identifying the level $k$ hyperplanes in $\widehat{L\g}_k^\ast$ in the
following fashion
\eqn\maptoconnections{\eqalign{
k \Lambda + L\g^\ast & \rightarrow {\frak A}\cr
k \Lambda + \mu & \mapsto {d\over d t} + {\mu\over k} \,.
}}
Further, acting on a highest weight module
$\H_\lambda \otimes \Spin$, the kernel of $\Dirac_\mu$ is localized on the
co-adjoint orbits, denoted $\frak{O}^{L\g}_\lambda$, of $\lambda +
\rho$, $\rho$ being the highest root. The kernels for each element in
$\frak{O}^{L\g}_\lambda$ are given by the image under the co-adjoint
action of the lowest-weight space of $\H_\lambda \otimes \Spin$, and
give rise to a twisted, equivariant vector bundle localized on $\frak{O}^{L\g}_\lambda$.

From this data one can now construct a twisted equivariant K-theory
class on $G$ by using \maptoconnections: namely, the co-adjoint action,
say for the level 1 hyperplane $\Lambda + L\g^\ast \subset
\widehat{L\g}$ is precisely given by the gauge action on a connection
\eqn\gaugeaction{
g \cdot (\mu, 1) =  (\hbox{ad}_{g} (\mu) - dg g^{-1},
1) \,,
}
for $g\in LG$.
Further, one can map a connection in ${\frak A}$ to an element in $G$ by utilizing the
holonomy map
\eqn\holmap{
\hbox{Hol}: \quad {\frak A} \rightarrow G \,.
}
Then the co-adjoint action, a.k.a. gauge action, in \gaugeaction\ maps
to
\eqn\holaction{
\hbox{Hol} (g \cdot (\mu, 1) ) = g(0) \hbox{Hol}(\mu) g(0)^{-1} \,. 
}
In this way we can map the Dirac family on $Lg^\ast$ to one on $G$,
and the equivariance with respect to $LG$ maps to an equivariance with
respect to the adjoint action of $G$ on itself. 
In summary, this construction thus
gives rise to a twisted K-theory class on $G$, equivariant under the
adjoint action of $G$.


\subsec{Dirac-family construction for the chiral ring}

Some related discussions for $H=T$ the maximal torus of $G$ has appeared in
\refs{\FHTloop}, relating $\tK_{G}(G)$ to $\tK_T(T)$. For our purposes,
we continue to assume only that $G$ is simply-connected and $G/H$ is
hermitian symmetric, thus ensuring the existence of ${\cal N}=2$
supersymmetry. 
Consider the affine Dirac operator $\Dirac \equiv \Dirac_{L\g/L\h}=\Dirac^- +\Dirac^+$ 
(\cf\ \affineDirac) acting on $\H_\lambda\otimes\Spin_{L\p}$
for $\H_{\lambda}$ an integrable highest weight representation of
$\widehat{L\g}_{k}$. This representation can be decomposed with respect to the subalgebra
$\widehat{L\h}$  
\eqn\gintohreps{
\H_\lambda = \bigoplus_{\nu\in L\h^\ast} M_{\lambda\nu} \bV_{\nu} \,,
}
where $\bV_{\nu}$ is a highest weight representation of $\widehat{L\h}$, $M_{\lambda\nu}$ being the state spaces of the coset theory.
Note that for the case of interest when $G/H$ is
hermitian symmetric, the Dirac operator simplifies drastically, as
$\p$ is commutative, so that $\Dirac = J\cdot \psi$. 

In \refs{\LVW} and \refs{\landweber} the kernel of
$\Dirac$ acting on $\H_\lambda\otimes\Spin_{L\p}$ is determined as
\eqn\dirackernel{
\left. \ker \Dirac_{L\g/L\h}\,  \right|_{\H_\lambda\otimes\Spin_{L\p}} =
\bigoplus_{(\lambda,\nu)\in {\frak C}} \bV_{\nu}\otimes\Spin_{L\p} \,,
} 
where\footnote{$^\sharp$}{This is in fact the condition for RR
groundstates, which we however shall use interchangeably with chiral
primaries, via spectral flow. More precisely the set of chiral
primaries is $(\mu, \nu)$ with $\sigma (\mu + \rho_\g) = \nu +
\rho_\g$. So we should denote by ${\frak C}$ the set of coset weights
satisfying this property.}
\eqn\Cdef{
{\frak C} = \left\{ (\mu,\nu)\in L\g^\ast \oplus L\h^\ast\,;\ \nu+ \hat\rho_{\h} = \sigma (\mu
+ \hat\rho_{\g})\,, \quad \hbox{for some } \sigma \in  \hat{W}_{\g}/\hat{W}_{\h}
\right\}  \,.
} 
The ring of chiral primaries $\Rcal_{cp}^{(G,H)}$ contains strictly speaking
only a subset of the chiral primaries in the kernel of
the Dirac operator. More precisely, it is generated by the primaries in $\ker(\Dirac)$,
which by \dirackernel\ is identified with the elements of ${\frak C}$,
modulo field identifications, which reside in the common centre $Z$ of $G$
and $H$, \ie, 
\eqn\chiralringiso{
\Rcal_{cp} = \left. \left\langle \Phi_{(\mu, \nu)}\, ; \ (\mu, \nu) \in
{\frak C} \right\rangle \right/ Z\,.
}

The following construction will show that for each element in ${\cal R}_{cp}$ there exists a twisted
$H$-equivariant K-theory class on $H$. 
Consider to begin with the
complex defined by $\Dirac_{L\g/L\h}$ acting on $\H_\lambda\otimes
\Spin_{L\p}$. 
Define a family of Dirac operators on $L\h^\ast$ as
\eqn\affineDiracfamily{\eqalign{
L\h^\ast  \quad \rightarrow & \quad  \End ((\H_\lambda \otimes
\Spin_\p, \Dirac_{L\g/L\h} ))  \cr
\mu \qquad \mapsto & \quad \Dirac^\mu_{L\h}\,,
}}
where we defined the (gauge-coupled) family of Dirac operators
\eqn\Diracmu{
\Dirac^\mu_{L\h} = \Dirac_{L\h} + i \psi(\mu)\,, \qquad \mu\in L\h^\ast\,.
}
Taking the kernel of $\Dirac^\mu_{L\h}$ on the complex $(\H_\lambda \otimes
\Spin_\p, \Dirac_{L\g/L\h} )$ amounts to decomposing the kernel of
$\Dirac_{L\g/L\h}$ as acting on $\H_\lambda\otimes\Spin$ into $\widehat{L\h}$ representations, meaning the
decomposition \dirackernel. So, this will result in a map from the
Verlinde ring of $\tK_{G}(G)$ to the one of $H$, \ie, 
\eqn\weylmap{\eqalign{
\tK_{G}(G) & \rightarrow {}^{\tau'}K_H(H) \cr
\Phi_{\lambda}^{\widehat{L\g}} & \mapsto \sum_{(\lambda,\nu) \in {\frak C}} \Phi_{\nu}^{\widehat{L\h}}\,.
}}
Note that the elements of the chiral ring can also
be viewed as semi-infinite Lie algebra cohomology elements, as discussed in \refs{\LVW}. Decomposing
$L{\frak p}= L{\frak p}^+ \oplus L {\frak p}^-$ such that $[{\frak p}^\pm,
{\frak p}^\pm] \subset {\frak p}^\pm$, the kernel of
$\Dirac_{L\g/L\h}$ acting on $\H_\lambda \otimes \Spin_{L\p}$ is 
\eqn\semicoho{
\left. \ker \Dirac_{L\g/L\h}\right|_{ \H_\lambda \otimes \Spin_{L\p} }
= H^\ast (L\p^\ast ; \H_\lambda) \,.
}
In this way one obtains a map from $V_{k}(G) \rightarrow
V_{k+h_\g^\vee - h_\h^\vee}(H)$, mapping precisely as in \weylmap.


\newsec{Discussions and Outlook}

In summary, we have defined a boundary ring ${\cal B}_k$ for ${\cal
N}=2$ coset models $(G, H)$ in terms of the twisted equivariant K-theory
$\tK_{H/Z}(G)$. The rank of ${\cal B}_k$ agrees with that of 
the chiral ring of the coset model, however the product structures
differ. 
Both rings are quotients of the representation ring of $H$,
with respect to the Verlinde ideals of $G$ and of $H$, at specific
levels, respectively. 

The present analysis is somewhat reminiscent of the theorem by
Freed-Hopkins-Teleman, which identifies the Verlinde algebra $V_k(G)$
with the twisted equivariant K-theory $\tK_{G}(G)$. Naively, one
might have anticipated an isomorphism between $\tK_{H/Z}(G)$ and the
chiral ring, which not only respects the structure as abelian groups,
but also the {\it products}. This is however not the case. 
More to the point, the present result suggests that
the K-theory associated to a particular sigma-model gives rise to an algebra
on classes of D-branes. This may well tie in with the algebra of BPS states
defined by Harvey and Moore \refs{\harveymoore, \bpsalgebra}.
The natural interpretation of FHT in this
light is, that for the topological $G/G$ coset model 
the D-brane charge relations obey the Verlinde algebra of $G$. 

It would be interesting to understand, what the precise relation
between ${\cal B}_k$ and the chiral ring is, \eg, if one can be
obtained as a deformation of the other, and possible relations to the
quantum cohomology ring may be interesting to explore. 
The boundary ring in the case of superminimal models, discussed in
section 3.2, turned out to be a deformation of the chiral ring (by
taking essentially only the highest degree component of the fusion
ideal). In this case, the boundary ring is a quantum deformed version
of the bulk chiral ring. Two immediate questions arise: firstly,
whether this holds for all KS coset models, \ie, the boundary ring is
given by a quantum (or otherwise) deformed bulk chiral ring and
secondly, whether this has any implications upon twisted K-theory.

One spin-off of our results is the agreement of the rank of the charge lattice of the
D-branes with the rank of the chiral ring.
This is a result that had been anticipated already in
\refs{\etkpaper} and suggests that the Cardy boundary states provide a
complete basis for the charge lattice. One can in fact define a
product on these Cardy boundary states
(and this will be discussed in detail in \refs{\stefantocome}), which should then agree with the product
in the boundary ring ${\cal B}_k$. 
Another interesting line of
thought would be to study the D-branes in the topologically twisted KS
models in this light. Such a worldsheet derivation of the boundary ring, 
whether in the full CFT or in the topologically
twisted model, will certainly substantiate the proposal put forward in
this paper, and may help elucidating the relation between boundary
conformal field theory and K-theory.


\vskip 1cm

\centerline{{\bf Acknowledgments}}
\pano

\noindent
I am grateful to Stefan Fredenhagen for important
discussions. Thanks also to
Volker Braun, Greg Moore and Constantin Teleman for interesting
comments, as well as Axel Kleinschmidt and Christian Stahn for {\it
mathematica}l advice related to the appendix. 
Hospitality of the IH\'ES during the ``Workshop avant Strings'' and of
DAMTP, Cambridge, is gratefully acknowledged. This work is partially supported by the European RTN Program HPRN-CT-2000-00148.


\appendix{A}{Computation of the ranks of the Verlinde algebras}

In this appendix we shall prove explicit formulae
for the rank of the fusion ring $d_k(G)$ for the groups $G$ appearing
in table 1. For the groups in question, the relations, of which for
fixed $k$ one needs to enumerate the non-negative integral solutions for $\{\Lambda^{(i)} \}$, are 
\eqn\ineqs{\eqalign{
A_n=SU(n+1)\, :& \qquad  \sum_{i=1}^n \Lambda^{(i)} \leq k \cr
D_n=SO(2n)  \, :& \qquad   \Lambda^{(1)} + \Lambda^{(n-1)} +\Lambda^{(n)}
                        + 2 \sum_{i=2}^{n-2} \Lambda^{(i)} \leq k  \cr
B_n=SO(2n+1) \, :&\qquad  \Lambda^{(1)} + \Lambda^{(n)} 
                        + 2 \sum_{i=2}^{n-1} \Lambda^{(i)} \leq k  \cr
C_n=Sp(2n)  \, : & \qquad   \sum_{i=1}^n \Lambda^{(i)} \leq k \cr
E_6 \,: & \qquad  \Lambda^{(1)}+ \Lambda^{(5)}+ 2
(\Lambda^{(2)}+\Lambda^{(4)}+\Lambda^{(6)})  + 3 \Lambda^{(3)}\leq k \cr 
E_7\, :& \qquad  \Lambda^{(6)} +2( \Lambda^{(1)} +\Lambda^{(5)} +
\Lambda^{(7)})+ 3(  \Lambda^{(2)}+ \Lambda^{(4)})+ 4 \Lambda^{(3)}\leq
k
\,.}}
In summary we obtain table 2, where
$k$ is the level of the affine algebra corresponding to $G$, $\kappa\in\Nop_0$ and $n\in\Nop$.
$$
\def\tbntry#1{\vbox to 23 pt{\vfill \hbox{#1}\vfill }}
\hbox{\vrule 
      \vbox{\hrule 
            \hbox{\vrule
                  \hbox to 40 pt{
                  \hfill\tbntry{$G$}\hfill }
                  \vrule 
                  \hbox to 40 pt{
                  \hfill\tbntry{ $k$ }\hfill }
                  \vrule 
                  \hbox to 120 pt{
                  \hfill\tbntry{ $d_k(G) $ }\hfill }
                  \vrule 
                }
            \hrule 
            \hbox{\vrule
                  \hbox to 40 pt{
                  \hfill\tbntry{$ A_n$}\hfill }
                  \vrule 
                  \hbox to 40 pt{
                  \hfill\tbntry{ $k$ }\hfill }
                  \vrule 
                  \hbox to 120 pt{
                  \hfill\tbntry{ ${n+k \choose k}$ }\hfill }
                  \vrule 
                }          
            \hrule 
            \hbox{\vrule
                  \hbox to 40 pt{
                  \hfill\tbntry{$B_{n+2}$}\hfill }
                  \vrule 
                  \hbox to 40 pt{
                  \hfill\tbntry{ $2\kappa$ }\hfill }
                  \vrule 
                  \hbox to 120 pt{
                  \hfill\tbntry{$
\ \, {n+\kappa+1\choose n+1} + 4 {n+\kappa+1\choose n+2}
 $ }\hfill }
                  \vrule 
                }                   
            \hrule 
            \hbox{\vrule
                  \hbox to 40 pt{
                  \hfill\tbntry{$B_{n+2} $}\hfill }
                  \vrule 
                  \hbox to 40 pt{
                  \hfill\tbntry{ $2\kappa+1$ }\hfill }
                  \vrule 
                  \hbox to 120 pt{
                  \hfill\tbntry{$
3{n+\kappa+1\choose n+1} + 4 {n+\kappa+1\choose n+2}
 $ }\hfill }
                  \vrule 
                }          
            \hrule 
            \hbox{\vrule
                  \hbox to 40 pt{
                  \hfill\tbntry{$ C_n$}\hfill }
                  \vrule 
                  \hbox to 40 pt{
                  \hfill\tbntry{ $k$ }\hfill }
                  \vrule 
                  \hbox to 120 pt{
                  \hfill\tbntry{ $ {n+k \choose k}$ }\hfill }
                  \vrule 
                }
            \hrule 
            \hbox{\vrule
                  \hbox to 40 pt{
                  \hfill\tbntry{$D_{n+3}$}\hfill }
                  \vrule 
                  \hbox to 40 pt{
                  \hfill\tbntry{ $2\kappa$ }\hfill }
                  \vrule 
                  \hbox to 120 pt{
                  \hfill\tbntry{ $ 
                  \ \, {n+\kappa+1\choose n+1} + 8 {n+\kappa+2\choose n+3} 
                  $ }\hfill }
                  \vrule 
                } 
            \hrule 
            \hbox{\vrule
                  \hbox to 40 pt{
                  \hfill\tbntry{$D_{n+3}$}\hfill }
                  \vrule 
                  \hbox to 40 pt{
                  \hfill\tbntry{ $2\kappa +1$ }\hfill }
                  \vrule 
                  \hbox to 120 pt{
                  \hfill\tbntry{ $ 
                  4{n+\kappa+2\choose n+2} + 8 {n+\kappa+2\choose n+3}
                  $ }\hfill }
                  \vrule 
                }
            \hrule 
            \hbox{\vrule
                  \hbox to 40 pt{
                  \hfill\tbntry{$E_6 $}\hfill }
                  \vrule 
                  \hbox to 40 pt{
                  \hfill\tbntry{ $k$ }\hfill }
                  \vrule 
                  \hbox to 120 pt{
                  \hfill\tbntry{ (A.14) }\hfill }
                  \vrule 
                }         
            \hrule 
            \hbox{\vrule
                  \hbox to 40 pt{
                  \hfill\tbntry{$E_7 $}\hfill }
                  \vrule 
                  \hbox to 40 pt{
                  \hfill\tbntry{ $k$ }\hfill }
                  \vrule 
                  \hbox to 120 pt{
                  \hfill\tbntry{ (A.17) }\hfill }
                  \vrule 
                }   
            \hrule     
        }
     }
$$
\centerline{
\hbox{{\bf Table 2:} {\it Ranks $d_k(G)$ of the Verlinde algebras $V_k(G)$.}}} 
\vskip 8pt

The remainder of this appendix will give the derivations of the relations in table 2. For $SU(n+1)$ and $Sp(2n)$
the argument is straight
forward. To prove the assertion in this case one can proceed by
induction upon $k$. The case $k=0$ is trivially satisfied. The
induction step follows by using
\eqn\binomi{
d_{k-1}(G) +{n+k -1 \choose k}= {n+k-1 \choose k-1} + {n+k -1 \choose
k} = {n+k \choose k }=  d_k(G)\,,
}
which implies the required formula, as the number of partitions of $k$
into $n$ parts is ${n+k-1\choose k}$.

Next, we consider $G=B_n= SO(2n+1)$. 
By using the result of $SU(2n+1)$
the rank of the fusion ring for $SO(2n+1)$ ($n\geq 3$) is
\eqn\dkalacs{
d_{k}(SO(2n+1)) = \sum_{\Lambda^{(1)}=0}^k
\sum_{\Lambda^{(n)}=0}^{k-\Lambda^{(1)}} {n-2 +
\left[{k-\Lambda^{(1)}-\Lambda^{(n)}\over 2}\right] \choose n-2} 
= \sum_{\Lambda=0}^k (\Lambda +1) {n-2 + \left[{k-\Lambda \over
2}\right]\choose n-2} \,.
}
Using
\eqn\binomiids{
\sum_{l=0}^k{n+l\choose n} = {n+1+k \choose n+1}\,,\qquad
\sum_{l=0}^{k} l {n+l\choose n} =(n+1) {n+k+1\choose n+2}\,,
}
one can sum these to obtain for even level $2k$
\eqn\soevenlevel{\eqalign{
&d_{2k}(SO(2(N+2)+1)) \cr
& =  -2 (N+1) {N+k+1\choose N+2} + (2k+1) {N+k+1 \choose N+1} 
-2 (N+1) {N+k \choose N+2} + 2k {N+k\choose N+1} \cr
&= -2 N {N+k+1\choose N+2} + (2k+1) {N+k+1 \choose N+1}
\,.
}}
This formula can be further simplified straight-forwardly by \eg,
expanding out one of the binomial coefficients
\eqn\soevenagain{
d_{2k}(SO(2(N+2)+1)) = {N+k+1 \choose N+1} + 4 {N+k+1\choose N+2} \,.
}
For odd level $2k+1$ the rank is computed by
\eqn\sooddlevel{\eqalign{
 d_{2k+1}(SO(2(N+2)+1))  &= 
(4\kappa +3) {N+\kappa +1\choose N+1} - 4(N+1) {N+\kappa +1\choose N+2}\cr
&= 3 {N+\kappa +1\choose N+1} + 4 {N+\kappa + 1\choose N+2}\,.
}}

Next consider $D_n$. We proceed analogously by computing
\eqn\dnseries{\eqalign{
d_k(SO(2n)) &=\sum_{\Lambda^{(1)}=0}^k\ 
\sum_{\Lambda^{(n)}=0}^{k-\Lambda^{(1)}} \ \sum_{\Lambda^{(n-1)}=0}^{k
-\Lambda^{(1)}-\Lambda^{(n)}} 
{n-3 +
\left[{k-\Lambda^{(1)}-\Lambda^{(n)} -\Lambda^{(n-1)}\over 2}\right] \choose n-3} \cr
&= 
\sum_{\Lambda=0}^k {(\Lambda+1) (\Lambda +2)\over 2} {n-3 + \left[{k-\Lambda \over 2}\right]\choose n-3} \,.
}}
In order to evaluate this, note that
\eqn\lsquaresum{\eqalign{
\sum_{l=0}^{k} l^2 {n+l\choose n} &=(n+1) (n+2) {n+k+2\choose n+3}
-(n+1)^2 {n+k+1\choose n+2}\cr
&= (n+1)(n+2){n+k+1\choose n+3}
+(n+1) {n+k+1\choose n+2}\,.
}}
Evaluation of the sum \dnseries\ yields for even level $k=2\kappa$
\eqn\dneven{\eqalign{
d_{2\kappa}(D_{n+3}) &=
\sum_{\Lambda=0}^k (2\kappa -2\Lambda +1)^2 {n+\Lambda \choose n}\cr
&= (2\kappa+1)^2 {n+\kappa +1\choose n+1} - 4 (n+1) {n+\kappa
+2\choose n+3 } - 4 \kappa (n+1) {n+\kappa +1\choose n+2}\cr
&= {n+\kappa+1\choose n+1} + 8 {n+\kappa+2\choose n+3} \,. 
}}
For odd level the rank is computed to be
\eqn\dneven{\eqalign{
&d_{2\kappa+1}(D_{n+3}) =
\sum_{\Lambda=0}^k 4 (\kappa -\Lambda +1)^2 {n+\Lambda \choose n}\cr
&\qquad = 4 (\kappa+1)^2 {n+\kappa+1\choose n+1}  -4(n+1) {n+\kappa+2 \choose n+3} -4 (\kappa+1)(n+1)
{n+\kappa+1\choose n+2} \cr
&= 4 {n+\kappa+2\choose n+2} + 8 {n+\kappa+2\choose n+3} \,. 
}}
The case of $E_6$ can be derived by using the result for $B_5$ 
\eqn\esix{
d_k(E_6) = \sum_{\Lambda=0}^k \left( \left[ {k-\Lambda\over 3}\right]
+1\right) (d_{\Lambda} (B_5)-d_{\Lambda-1}(B_5))\,.
}
There are three cases to be considered: $k=3\kappa$, $k=3\kappa+1$ and
$k=3\kappa+2$, for $\kappa\in\Nop_0$. In each of these cases we
obtain the following sums
\eqnn\esixone
\eqnn\esixtwo
$$
\eqalignno{
d_{3\kappa+ s}(E_6) & = \sum_{\Lambda=0}^\kappa (\kappa-\Lambda +1)
\left(d_{3\Lambda+s}(B_5) - d_{3\Lambda-3+s}(B_5)\right)& \esixone\cr
& =\left( \sum_{\Lambda=0}^\kappa d_{3\Lambda +s}(B_5) \right)-(\kappa +1) d_{s-3}(B_5) \ ,\qquad s=0,1,2
\,. &\esixtwo
}
$$
Note, that $d_\kappa (G)=0$ for $\kappa <0$.

For $E_7$ one again proceeds stepwise. For fixed value of 
$k_4=\Lambda^{(6)} +2( \Lambda^{(1)} +\Lambda^{(5)} +
\Lambda^{(7)})$ one has 
\eqn\gfour{
\sum_{\Lambda^{(6)}=0}^{k_4} {3+ \left[{k_4-\Lambda^{(6)}\over 2}\right]\choose
3}\,,
}
and for fixed $k_6=\Lambda^{(6)} +2( \Lambda^{(1)} +\Lambda^{(5)} +
\Lambda^{(7)}) + 3 (\Lambda^{(2)} + + \Lambda^{(4)})$ 
\eqn\gsix{
n_{k_6}=
\sum_{\Lambda^{(4)}=0}^{k_6} \,
\sum_{\Lambda^{(2)}=0}^{k_6-\Lambda^{(4)}}
\sum_{\Lambda^{(6)}=0}^{\left[{k_6-\Lambda^{(2)}-\Lambda^{(4)}\over 3}\right]} 
\left(\matrix{
& 3+ \left[{\left[{k_6-\Lambda^{(2)}-\Lambda^{(4)}\over 3}\right]
 -\Lambda^{(6)}\over 2}\right]\cr
& 3}\right)
}
Iterating this procedure, we arrive at
\eqn\eseven{
d_k(E_7) =
\sum_{\Lambda^{(3)}=0}^k  n_{\Lambda^{(3)}} \,.
}
In summary we obtain table 2. 

Note that these multiplicity formulae can also be extracted from the generating function
\eqn\xelagenfu{\eqalign{
A_{(a_1, \cdots, a_n)}(q) 
&=(1-q)^{-1} \prod_{i=1}^n (1-q^{a_i})^{-1} \cr
&= \sum_{k=0}^{\infty} d_k(G) q^k \,,
}}
which counts the non-negative integer solutions $\{\Lambda^{(i)}\}$ to 
\eqn\generalineq{
a_1 \Lambda^{(1)} + \cdots +a_n \Lambda^{(n)} \leq k\,,
}
where the choice of group $G$ determines the coefficients
$a_i\in\Nop_0$ as \eg\ in \ineqs.


\listrefs

\bye